\documentstyle[twocolumn,aps,epsfig]{revtex}

\begin{document}
\title{Quasi-stable black holes at the large hadron collider}

\author{Sabine Hossenfelder, Stefan Hofmann, Marcus Bleicher, Horst St\"ocker}

\address{Institut f\"ur Theoretische Physik\\ J. W. Goethe Universit\"at\\60054 Frankfurt am Main, Germany}

\maketitle

\noindent
\begin{abstract}
We address the production of black holes at LHC
and their time evolution in space times 
with compactified space like extra dimensions.
It is shown that black holes with life times of 
several hundred fm/c can be produced at LHC. 
The possibility of quasi-stable remnants is discussed.
\vspace{1cm}
\end{abstract}

An outstanding problem in physics is to understand the ratio
between the electroweak scale $m_W=10^3$~GeV
and the four-dimensional Planck scale $m_p=10^{19}$~GeV.
Proposals that address this so called
hierarchy problem within the context of brane world scenarios
have emerged recently \cite{add}.
In these scenarios the Standard Model of particle physics
is localized on a three dimensional brane in a higher
dimensional space. This raises the exciting possibility
that the fundamental Planck scale $M_f$ can be as low as $m_W$.
As a consequence, future high energy colliders like LHC
could probe the scale of quantum gravity with its exciting new 
phenomena: A possible end of small
distance physics as been investigated by Giddings and Thomas\cite{gid} 
while Dimopoulos 
and Landsberg opened a new road to study black holes with their 
work on the production of black holes in high 
energetic interactions \cite{dim}.
In this letter we investigate TeV scale gravity 
associated with black hole production and evaporation
at LHC and beyond. For a discussion of black hole production at Tevatron and 
from cosmic rays, the reader is referred to 
Refs. \cite{Bleicher:2001kh,Ringwald:2001vk,Anchordoqui:2001cg}.

One scenario for realizing TeV scale gravity is a
brane world in which the Standard Model particles
including gauge degrees of freedom reside on a 
3-brane within a flat compact space of volume
$V_d$, where $d$ is the number of compactified spatial
extra dimensions with radius $L$.
Gravity propagates in both the compact and non-compact
dimensions.

Let us first characterize black holes
in space times with compactified space-like
extra dimensions \cite{adm}.
We can consider two cases:
\begin{enumerate}
\item The size of the black hole given by its
Schwarz\-schild radius $R_H$ is $\gg L$.
\item If $R_H \ll L$ the topology of the horizon
is spherical in $3+d$ space like dimensions.
\end{enumerate}
The mass of a black hole with $R_H \approx L$
in $D=4$ is called the critical mass
$M_c \approx m_p L/l_p$ and $1/l_p=m_p$.
Since 
\begin{equation}
L\approx
\left(1 {\rm TeV}/M_f \right)^{1+\frac{2}{d}}
10^{\frac{31}{d}-16} \; {\rm mm}
\end{equation} 
$M_c$ is typically of the order of the Earth mass. 
Since we are interested in black
holes produced in parton-parton collisions
with a maximum c.o.m. energy of $\sqrt{s}=14$~TeV,
these black holes have $R_H \ll L$ 
and belong to the second case.

Spherically symmetric solutions describing black holes
in $D=4+d$ dimensions have been obtained\cite{my}
by making the ansatz
\begin{equation}
{\rm d}s^2 =
- {\rm e}^{2\phi(r)} {\rm d}t^2
+ {\rm e}^{2\Lambda(r)} {\rm d}r^2
+ r^2 {\rm d}\Omega_{(2+d)}
\; ,
\end{equation}
with ${\rm d}\Omega_{(2+d)}$ denoting the
surface element of a unit $3+d$-sphere.
Solving the field equations $R_{\mu\nu} = 0$
gives
\begin{equation}
{\rm e}^{2\phi(r)} = {\rm e}^{-2\Lambda(r)}
=
1-\left(\frac{C}{r}\right)^{1+d},
\end{equation}
with $C$ being a constant of integration.  
We identify $C$ by the requirement
that for $r\gg L$ the force derived from the potential 
in a space time with $d$ compactified extra dimensions 
\begin{equation}
V(r)
=
\frac{1}{(1+d)} 
\left( \frac{1}{M_f}\right)^{1+d} \frac{M}{M_f}\frac{1}{L^d} \frac{1}{r}
\end{equation}
equals the force derived from the usual $4$-dimensional Newton potential. 
Note, the mass $M$ of the black hole is defined
by
\begin{equation}
M
=
\int {\rm d}^{3+d} x
\; T_{00}
\end{equation}
with $T_{\mu\nu}$ denoting the energy momentum tensor
which acts as a source term in the Poisson equation
for a slightly perturbed metric in
$3+d$ dimensional space time \cite{my,prep}.
In this way the horizon radius is obtained as
\begin{equation}
R_H^{1+d}=
\frac{2}{d+1} 
\left(\frac{1}{M_f}\right)^{1+d} \; \frac{M}{M_f}
\end{equation}
with $M$ denoting the black hole mass.

Let us now investigate the production rate
of these black holes at LHC. 
Note that  we neglect complications due to the finite angular
momentum and assume non-spinning black holes in the formation 
and evaporation process\footnote{In fact finite angular momentum
decreases the production probability by $\approx 50$\% 
at LHC energies\cite{Anchordoqui:2001cg,hossi2002}.}.  
Consider two partons moving in opposite
directions. If the partons center of mass energy
$\sqrt{\hat s}$ reaches the fundamental
Planck scale $M_f\sim 1$~TeV
and if the impact parameter is less than $R_H$,
a black hole with Mass $M\approx \sqrt{\hat s}$
can be produced.
The total cross section for such a process
can be estimated on geometrical grounds \cite{Thorne:1972ji,bf}
and is of order $\sigma(M)\approx \pi R_H^2$.
This expression contains only the fundamental
Planck scale as a coupling constant.
Note that the given  
classical estimate of the black hole production 
cross section has been under debate\cite{Voloshin:2001fe,Giddings:2001ih}.
{{Investigations by \cite{Solodukhin:2002ui} 
justify the use of 
the classical limit \footnote{%
Even if this additional suppression would be present, the maximal 
suppression in the cross section is by a 
factor $10^{-1}$\cite{Rizzo:2001dk}.}. Further, the recent works done by Jevecki
and Thaler\cite{Thaler} and those done by Eardley and Giddings\cite{Giddings}
support this estimate.}}
Thus, we proceed with the classical approximation. 
Setting $M_f\sim 1$TeV and $d=2$ one finds 
$\sigma \approx 1$~TeV$^{-2}\approx 400$~pb.
However, we have to take into account
that in a pp-collision 
each parton carries only a fraction of the
total c.o.m. energy. The relevant quantity
is therefore 
the Feynman $x$ distribution
of black holes at LHC for masses
$M\in[M^{-},M^{+}]$ 
given by
\begin{eqnarray}
\frac{d\sigma}{d x_F} &=& \sum\limits_{p_1,p_2} 
\int\limits_{M^{-}}^{M^{+}} dy \\ 
&&\frac{2 y}{x_2 s} f_1(x_1, Q^2) f_2(x_2, Q^2) 
\sigma(y,d)\; ,
\end{eqnarray}
with $x_F=x_2-x_1$ and the restriction 
$x_1 x_2 s=M^2$.
We used the CTEQ4 \cite{lai1} parton distribution functions $f_1$, $f_2$ 
with $Q^2=M^2$. 
All kinematic combinations of partons from 
projectile $p_1$ and target $p_2$ are summed over. 
\begin{figure}[h]
\vskip 0mm
\vspace{-1.3cm}
\centerline{\psfig{figure=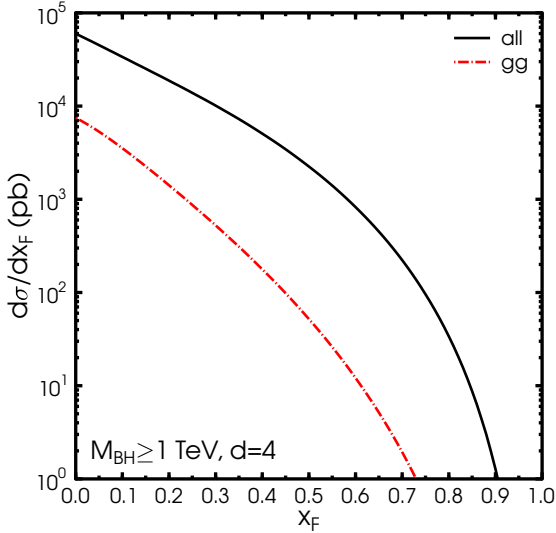,width=3.5in}}
\vskip -5mm
\caption{Feynman $x$ distribution of black holes  
with $M\ge 1$~TeV produced in pp interactions at LHC with 4 
compactified spatial extra dimensions. 
\label{dndy}}
\end{figure}
Fig. \ref{dndy} depicts the momentum distribution of produced 
black holes in pp interactions at $\sqrt s = 14$~TeV.
The Feynman $x$ distribution scales with
the black hole mass like $M^{2(2+d)/(1+d)}$.
As a consequence, 
black holes of lowest masses ($\approx 1$~TeV) receive 
a major contribution from $gg$ scattering, while heavier 
black holes are 
formed in scattering processes of 
quarks. 
Since for masses below $10$~TeV
heavy quarks give a vanishing contribution 
to the black hole production cross section,
those black holes are primarily formed in
scattering processes of
$up$ and $down$ quarks. 

Let us now investigate the evaporation of 
black holes with $R_H \ll L$ and 
study the influence of compact extra dimensions 
on the emitted quanta.
In the framework of black hole thermodynamics 
the entropie $S$ of a black hole is given by its surface
area. In the case under consideration
$S\sim M_{f}^{(2+d)} \Omega_{(d+3)} \; R_H^{2+d}$ \cite{prep}
 with $\Omega_{(d+3)}$ the surface of the unit $d+3$-sphere
\begin{eqnarray}
\Omega_{(d+3)} = \frac{2 \pi^{\frac{d+3}{2}}}{\Gamma({\frac{d+3}{2}})}\;\;.
\end{eqnarray}
The single particle spectrum of the emitted
quanta is then
\begin{equation}
n(\omega) 
= 
\frac{{\rm exp}[S(M-\omega)]}{{\rm exp}[S(M)]}\quad.
\end{equation}
\begin{figure}[h]
\vskip 0mm
\vspace{0cm}
\centerline{\psfig{figure=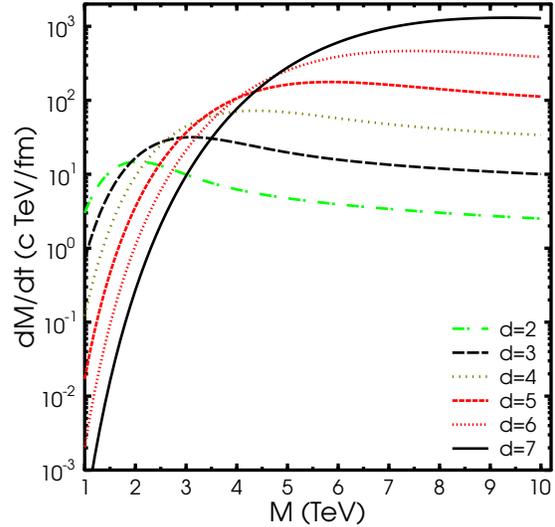,width=3.5in}}
\vskip 2mm
\caption{Decay rate in TeV$c$/fm as a function of the 
initial mass of the black hole. 
Different line styles correspond to different numbers of 
extra dimensions $d$.  
\label{mdot}}
\end{figure}
It has been claimed that it may not be possible
to observe the emission spectrum directly,
since most of the energy is radiated in
Kaluza-Klein modes. However, from the
higher dimensional perspective this seems
to be incorrect and most of the energy
goes into modes on the brane.
In the following we assume that most
of the emitted quanta will be localized 
on our 3-brane \cite{ehm}.

Summing over all possible multi-particle spectra 
we obtain the black holes evaporation 
rate $\dot{M}$ through the 
Schwarzschild surface ${\cal A}_D$ in $D$ space-time dimensions,
\begin{equation}
\dot{M} = - {\cal A}_{D} \frac{\Omega_{(d+3)}}{(2 \pi)^{d+3}}
\int\limits_0^{M} 
{\rm d}\omega\; \sum\limits_{j=1}^{(M/\omega)} 
\omega^{D-1} n(j\omega) \quad.
\label{upsmdot}
\end{equation}
Neglecting finite size effects 
Eq.(\ref{upsmdot}) becomes
\begin{eqnarray}
\dot{M} &=& {\cal A}_D \frac{\Omega_{(d+3)}}{(2 \pi)^{d+3}}\; 
{\rm e}^{-S(M)} \sum\limits_{j=1}^{\infty}
\left(\frac{1}{j}\right)^D \times \nonumber\\
&& \int\limits_{M}^{(1-j)M} {\rm d}x\; 
(M-x)^{D-1} {\rm e}^{S(x)} \Theta(x)\quad, 
\end{eqnarray}
with $x= M- j\omega$, 
denoting the energy of the black hole after emitting $j$ 
quanta of energy $\omega$. 
Thus, ignoring finite size effects 
we are lead to the 
interpretation that the black hole emits only a single 
quanta per energy interval.
We finally arrive at
\begin{eqnarray}
\dot{M} &=& {\cal A}_D \zeta (D) \frac{\Omega_{(d+3)}}{(2 \pi)^{d+3}} {\rm e}^{-S(M)} \times \nonumber\\
&& \int\limits_0^M {\rm d} x \; 
(M-x)^{D-1} {\rm e}^{S(x)} \; . \label{mdoteq}
\end{eqnarray}
Fig. \ref{mdot} shows the decay rate (\ref{mdoteq})
in TeV$c$/fm as a function of the initial mass of the black hole. 
Since the Temperature $T_h$ of the black hole 
decreases like $M^{-1/(1+d)}$ it is evident
that extra dimensions help stabilizing
the black hole, too. One should note that 
the mass decay law presented here is more complicated than 
the one derived in \cite{Page:df}. This is due to the micro
canonical treatment used here compared to the
grand canonical approach given in \cite{Page:df}. {{The grand canonical 
approach is suited only in settings when the energy of the emitted 
particle can be neglected as might be the case for astrophysical black holes.}}
For a detailed comparison between the micro canonical and the canonical
approach the reader is referred to \cite{Casadio:2001dc,prep}.

\begin{figure}[h]
\vskip 0mm
\vspace{0cm}
\centerline{\psfig{figure=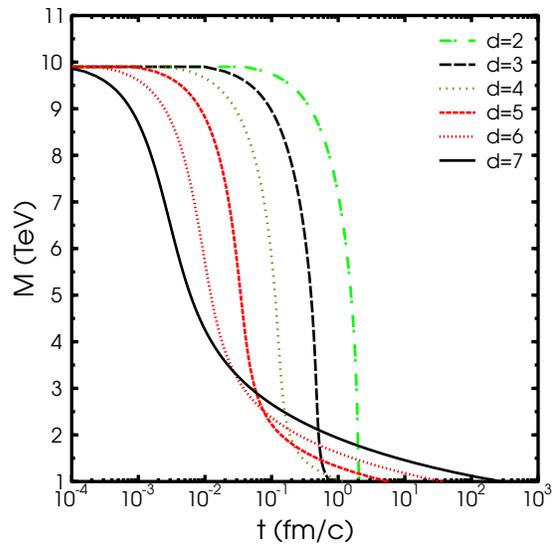,width=3.5in}}
\vskip 2mm
\caption{Time evolution of a black hole. 
Different line styles correspond to different numbers of 
extra dimensions $d$.  
\label{mt2}}
\end{figure}

From (\ref{mdoteq}) we calculate the time evolution
of a black hole with given Mass $M$. The result is 
depicted in Fig. \ref{mt2} for different numbers
of compactified space like extra dimensions. 
As can be seen again, extra dimensions lead to an increase 
in lifetime of black holes. 
The calculation shows that a black hole with $M\sim$ 10~TeV 
at least exist for $100$~fm/c for $d>5$ and is sensitive to the
number of extra dimensions. If the black hole has 
been created at large $x_F$ its apparent 
lifetime in the center of mass frame may even be larger by a 
factor seven due to relativistic time delay.
Afterwards
the mass of the black hole drops below the fundamental
Planck scale $M_f$. The quantum physics at this scale
is unknown and therefore the fate of the 
extended black object. 
However, statistical mechanics may still be valid.
If this would be the case it seems that after
dropping below $M_f$ a quasi-stable remnant remains.

In conclusion, we have predicted the momentum distribution
of black holes in space times with large and compact 
extra dimensions. Using the micro canonical ensemble
we calculated the decay rate of black holes
in this space time neglecting finite size effects.
If statistical mechanics is still valid below
the fundamental Planck scale $M_f$, the black holes
may be quasi-stable. In the minimal scenario 
($M_f\sim$ 10~TeV, $d>5$) the lifetime is at least
$100$~fm/c.   

The authors want to thank L. Gerland and  D.~J.~Schwarz 
for fruitful and stimulating discussions.
This work was supported in parts by BMBF, DFG, and GSI.

Note added in proof:
In previous proceedings (J.Phys.G28:1657,2002; hep-ph/0111052 and
Proceedings of the  XL International Winter meeting on Nuclear Physics,
p.58), we have published figures for the lifetimes and
evaporation rates of black holes with an omitted pre-factor that has
been included here.

\end{document}